\documentclass[aps,prl,twocolumn,superscriptaddress,longbibliography]{revtex4-1}
\usepackage{amsmath}
\usepackage{hyperref}
\hypersetup{
	colorlinks=true,
	linkcolor=blue,
	filecolor=blue,
	citecolor = blue,      
	urlcolor=blue,
}

\usepackage{graphicx}
\usepackage{dcolumn}
\usepackage{bm}
\usepackage{natbib}
\usepackage[utf8]{inputenc}
\usepackage{units}
\usepackage{amsmath}
\usepackage{amssymb}
\usepackage{graphicx}
\usepackage{bm}
\usepackage{multirow,microtype,color,relsize,ulem}

\usepackage[utf8]{inputenc}
\usepackage[T1]{fontenc}
\usepackage{mathptmx}
\usepackage{amsfonts}
\usepackage{amsmath}

\begin{document}

\title{Quantum enhanced SU(1,1) matter wave interferometry in a ring cavity}
\author{Ivor Kre\v{s}i\'{c}} \email{ivor.kresic@tuwien.ac.at} 
\affiliation{Institute for Theoretical Physics, Vienna University of Technology (TU Wien), Vienna, A–1040, Austria}
\affiliation{Centre for Advanced Laser Techniques, Institute of Physics, Bijeni\v{c}ka cesta 46, 10000, Zagreb, Croatia}
\author{Thorsten Ackemann} 
\affiliation{SUPA and Department of Physics, University of Strathclyde, Glasgow G4 0NG, Scotland, UK}

\date{\today}

\begin{abstract}
Quantum squeezed states offer metrological enhancement as compared to their classical counterparts. Here, we devise and numerically explore a novel method for performing SU(1,1) interferometry beyond the standard quantum limit, using quasi-cyclic nonlinear wave mixing dynamics of ultracold atoms in a ring cavity. The method is based on generating quantum correlations between many atoms via photon mediated optomechanical interaction. Timescales of the interferometer operation are here given by the inverse of photonic recoil frequency, and are orders of magnitude shorter than the timescales of collisional spin-mixing based interferometers. Such shorter timescales should enable not only faster measurement cycles, but also lower atomic losses from the trap during measurement, which may lead to significant quantum metrological gain of matter wave interferometry in state of the art cavity setups.

\end{abstract}

\maketitle
The study of light mediated atomic self-organization has advanced greatly since the pioneering experiments in hot alkali vapours \cite{grynberg_observation_1988,pender90,grynberg94,ackemann94}. With the maturation of laser cooling and trapping techniques, self-organizing instabilities in laser driven ultracold atoms have subsequently been researched in a wide variety of feedback schemes, establishing a rich subfield of atomic physics \cite{ritsch_cold_2013,mivehvar_cavity_2021,domokos_collective_2002,black_observation_2003,baumann_dicke_2010,nagy_dicke-model_2010,gopalakrishnan_atom-light_2010,greenberg_bunching-induced_2011,labeyrie_optomechanical_2014,robb_quantum_2015,schmittberger_spontaneous_2016,ostermann_spontaneous_2016,ballantine_meissner-like_2017,leonard_supersolid_2017,kollar_supermode-density-wave-polariton_2017,kresic18,vaidya_tunable-range_2018,guo_sign-changing_2019,guo_emergent_2019,schuster2020,baio_multiple_2021,guo_optical_2021,ackemann_self-organization_2021}. 

The earlier works on the quantum aspects of ultracold atom-cavity interaction have concentrated on studying steady state quantum correlations between light and atoms \cite{mekhov_probing_2007,nagy_nonlinear_2009,nagy_dicke-model_2010,nagy_critical_2011,elliott_multipartite_2015,ostermann_unraveling_2020,ivanov_feedback-induced_2020}. Recently, the generation of correlated atomic pairs via cavity light-mediated interaction and self-organization, has also come into focus \cite{davis19,finger23,kresic2023}, inspired by the earlier work on photon quantum correlations in optical parametric amplifiers and self-organized optical structures in nonlinear crystals \cite{louisell61,caves84,yurke86,lugiato_quantum_1992,grynberg_quantum_1993,marzoli97}. These recent works shift the attention from light-atom entanglement, which was studied in \cite{mekhov_probing_2007,nagy_nonlinear_2009,nagy_dicke-model_2010,nagy_critical_2011,elliott_multipartite_2015,ostermann_unraveling_2020,ivanov_feedback-induced_2020}, towards light-mediated atom-atom entanglement generation in a cavity.

The importance of quantum entangled states in quantum technologies lies in their ability to speed up a number of computational \cite{deutsch92} and metrological tasks \cite{giovannetti_quantum_2006,clerk_introduction_2010}. Regarding the latter, quantum enhanced measurement schemes with internal atomic degrees of freedom~\cite{duan2000,pu2000,sorensen_many-particle_2001,gross_nonlinear_2010,leroux_implementation_2010,schleier-smith_squeezing_2010,gross_spin_2012,hosten_measurement_2016,luo2017,pezze_quantum_2018,davis19,anjun2021}, and also the motional ones~\cite{salvi2018,gietka2019,shankar2019,anders2021,greve2022entanglement,finger23}, have been explored recently.

In this Article, we start with a U(1) symmetric Hamiltonian describing optomechanical stripe ordering in a Bose-Einstein condensate (BEC) placed inside a transversely pumped ring cavity, and show that its transient dynamics near pump threshold can be described by a SU(1,1) Hamiltonian \cite{law98,pu99,chang05,zhang05,gerving12}. By applying the insight from \cite{liu22} that cyclic dynamics can lead to effective time reversal in such a quantum system, we numerically demonstrate quantum enhanced SU(1,1) matter wave interferometry with the ring cavity scheme. Interferometric estimation of the phase shift using measurements of mean value and variance of the atomic on-axis momentum mode number operator \cite{liu22}, allows for precision measurements of the optical transition recoil frequency. Combining this quantity, with the result of a corresponding transition wavelength measurement, can be used to determine the fine structure constant \cite{weiss93,bouchendira11,parker18}, and inertial mass at microscopic scales \cite{bongs19}.

In contrast to the previously studied schemes for nonlinear SU(1,1) spin state interferometry with Bose-Einstein condensates \cite{chen15,linnemann16,liu22}, our proposal employs atoms with a single ground state (spin-0), for matter wave (motional state) interferometry. Due to relative simplicity of the setup, these results highlight the potential of employing ultracold atomic self-organization for quantum technologies.

\begin{figure}[!t]
\centering
\includegraphics[clip,width=\columnwidth]{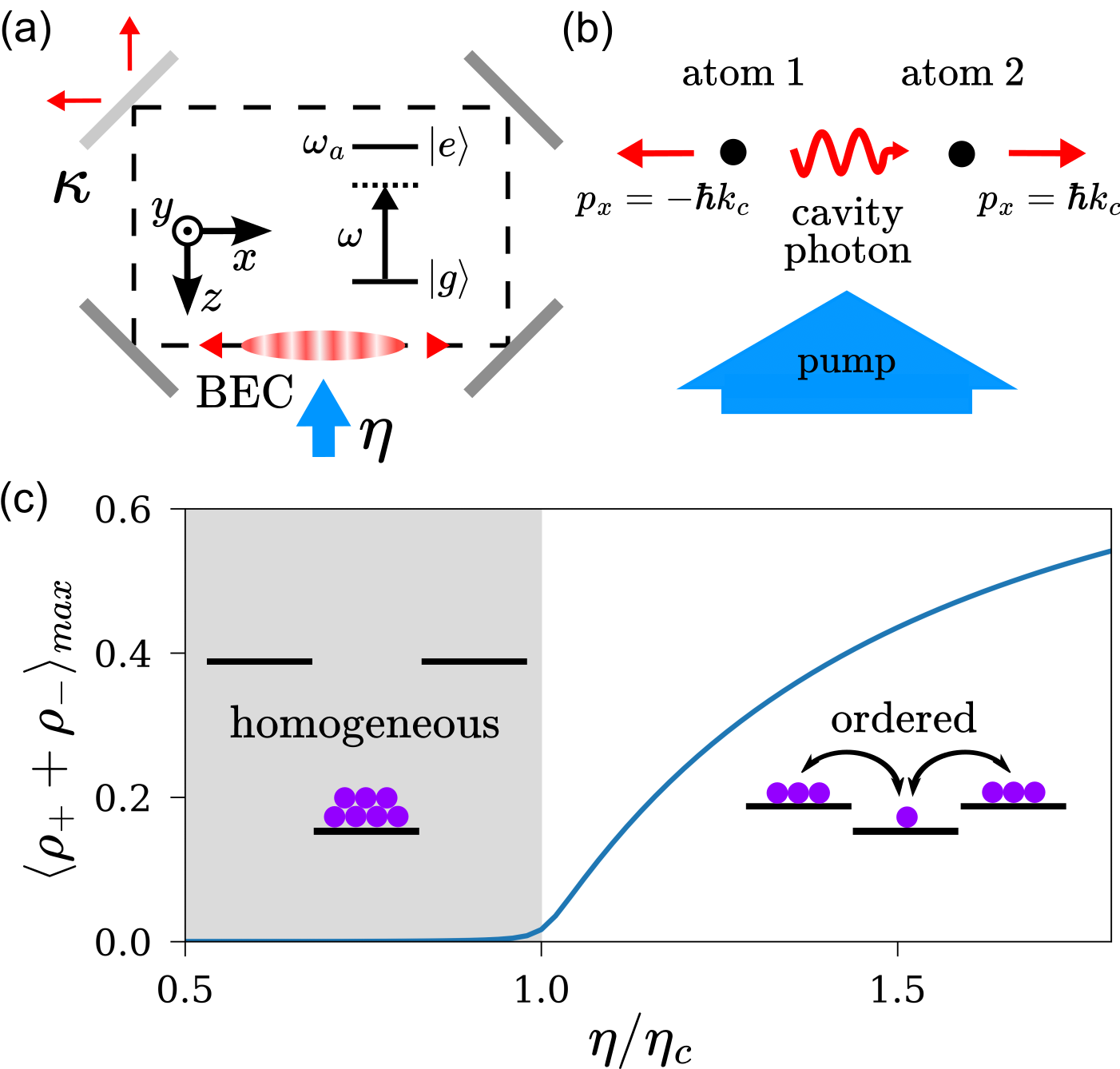}
\caption{Principle of entanglement generation in a Bose-Einstein condensate (BEC) placed inside a transversely pumped ring cavity. (a) Self-organization of laser pumped atoms in a ring cavity ($\eta$ - pump rate) with photon leakage rate $\kappa$. The two-level atomic optical transition with frequency $\omega_a$ is driven by a far off-resonant laser beam of frequency $\omega$. (b) An atom in the condensate gets a momentum kick of $-\hbar k_c$ ($\hbar k_c$) by scattering a drive photon with wavenumber $k= k_c$ into the initially empty counterclockwise (clockwise) cavity mode with wavenumber $k_c$. A correlated atom with $\hbar k_c$ ($-\hbar k_c$) can then be created if this cavity photon does not decay out of the cavity but scatters back into the driving field. (c) Scan of largest $\langle\rho_++\rho_-\rangle=\langle N_++N_-\rangle /N$ attained during unitary evolution (see text), against pump rate $\eta$. For $\eta<\eta_c$, the energy cost prohibits the excitation of atoms into $p_x=\pm\hbar k_c$ states, leading to a homogeneous BEC. In contrast, macroscopic populations in the $p_x=\pm\hbar k_c$ states occur when $\eta>\eta_c$, leading to striped order. Parameters: $N=1000$, $\bar{\Delta}_c=-1$ GHz, $\omega_R=2\pi\times 14.5$ kHz. }\label{Fig:1}
\end{figure}

The setup is shown in Fig. \ref{Fig:1}a). It consists of a prolate shaped Bose-Einstein condensate (BEC) held inside a ring cavity, and pumped along the $-z$ direction by a coherent field with pump rate $\eta$, frequency $\omega$ and wavenumber $k$. As in earlier work on transversely pumped cavities \cite{nagy_dicke-model_2010,domokos_collective_2002}, we study the 1D situation, where the recoil along the $z$ axis is neglected due to a trap confining the atoms along $y$- and $z$-axes \cite{mivehvar18}. A similar setup has been experimentally implemented in \cite{schuster2020}. Contrary to the similar recently utilized mechanisms for entanglement generation using atoms with multilevel transitions \cite{davis19,finger23}, the situation studied here relies on atoms and light interacting via a two-level optical transition.

The free space photon scattering can be greatly
suppressed in atom-cavity systems with collective
strong coupling \cite{brennecke08,ritsch_cold_2013,finger23}, such that the atom-light interaction is well described by taking into account only the intracavity photon modes. For light far-detuned from the atomic transition, the excited state can be adiabatically eliminated, leading to a Hamiltonian describing optomechanical interaction. Using the three optomechanical mode approximation, which is a good description at $\eta$ values near threshold \cite{mivehvar18}, the atomic motion can be described by a zero-order mode with $p_x=0$, and left- and right-moving modes with $p_x=\mp\hbar k_c$, with annihilation operators $b_j$ where $j=0,+,-$, and the field operator given by:
\begin{align}\label{eq:atomfield}
\psi(x)=\frac{1}{\sqrt{V}}\left(b_0+b_+e^{ik_cx}+b_-e^{-ik_cx}\right),
\end{align}
with $V$ being the volume of the system and $k_c$ the wavenumber of the ring cavity modes. As the pump-cavity detunings we use are many orders of magnitude smaller than the cavity frequency, in the above we have taken $k=k_c$. We here assume that relevant system dynamics is significantly faster than the cloud expansion in the harmonic trap, such that the description of the cloud as a quantum degenerate gas with three modes is valid throughout \cite{finger23}.

Adiabatically eliminating the photonic fields, the unitary evolution of the atomic degrees of freedom is determined by the effective Hamiltonian (see Appendix A for a detailed derivation):
\begin{align}\label{eq:startingham}
H_{c}& = \frac{g_c}{2N}[2b_+^\dagger b_-^\dagger b_0 b_0+2b_0^\dagger b_0^\dagger b_+ b_-\\
+& (2N_0-1)(N_++N_-)]-qN_0,
\end{align}
where $g_c=2N\bar{\Delta}_c\eta^2/(\bar{\Delta}_c^2+\kappa^2)=-\omega_R \eta^2/(2\eta_c^2)$, $q=\omega_R+g_c/N$, with $\eta_c=\sqrt{-\omega_R(\bar{\Delta}_c^2+\kappa^2)/(4\bar{\Delta}_cN)}$, and $N_j=b^\dagger _j b_j$. Here $N=N_0+N_++N_-$ is the total number of atoms, kept constant in the simulations presented, $\bar{\Delta}_c$ is the detuning of the pump laser from the cavity mode, $\kappa$ is the cavity photon decay rate, $\omega_R=\hbar^2 k_c^2/(2m)$ is the photon recoil frequency, and $\eta_c$ is the threshold pump rate for self-organization.

The first two terms in Eq. (\ref{eq:startingham}) describe the creation and destruction (mixing) of correlated atom pairs with opposite momenta $\pm\hbar k_c$, from the initial polar state $|N\rangle_0|0\rangle_+|0\rangle_-$. The third term describes the energy shift caused by the photon-mediated interatomic elastic collision processes that do not produce correlated atom pairs. The fourth term describes the energy shift of the momentum ordered modes with $p_x=\pm\hbar k_c$ with respect to the homogeneous mode $p_x=0$. For $g_c<0$, the system undergoes self-organization above a quantum critical point at $q=2|g_c|$ \cite{sadler06,luo2017,liu22}, where it becomes more energetically favorable to populate the $p_x=\pm \hbar k_c$ states via the mixing terms, as illustrated in Fig. \ref{Fig:1}c). For large $N$, $q=2|g_c|$ corresponds to the semiclassical threshold condition $\eta=\eta_c$. In the semiclassical picture, self-organization in atomic density occurs above threshold due to an optical lattice arising from interference of superradiantly scattered light in the co- and counter-propagating cavity modes \cite{mivehvar18,schuster2020}.

\begin{figure}[!t]
\centering
\includegraphics[clip,width=\columnwidth]{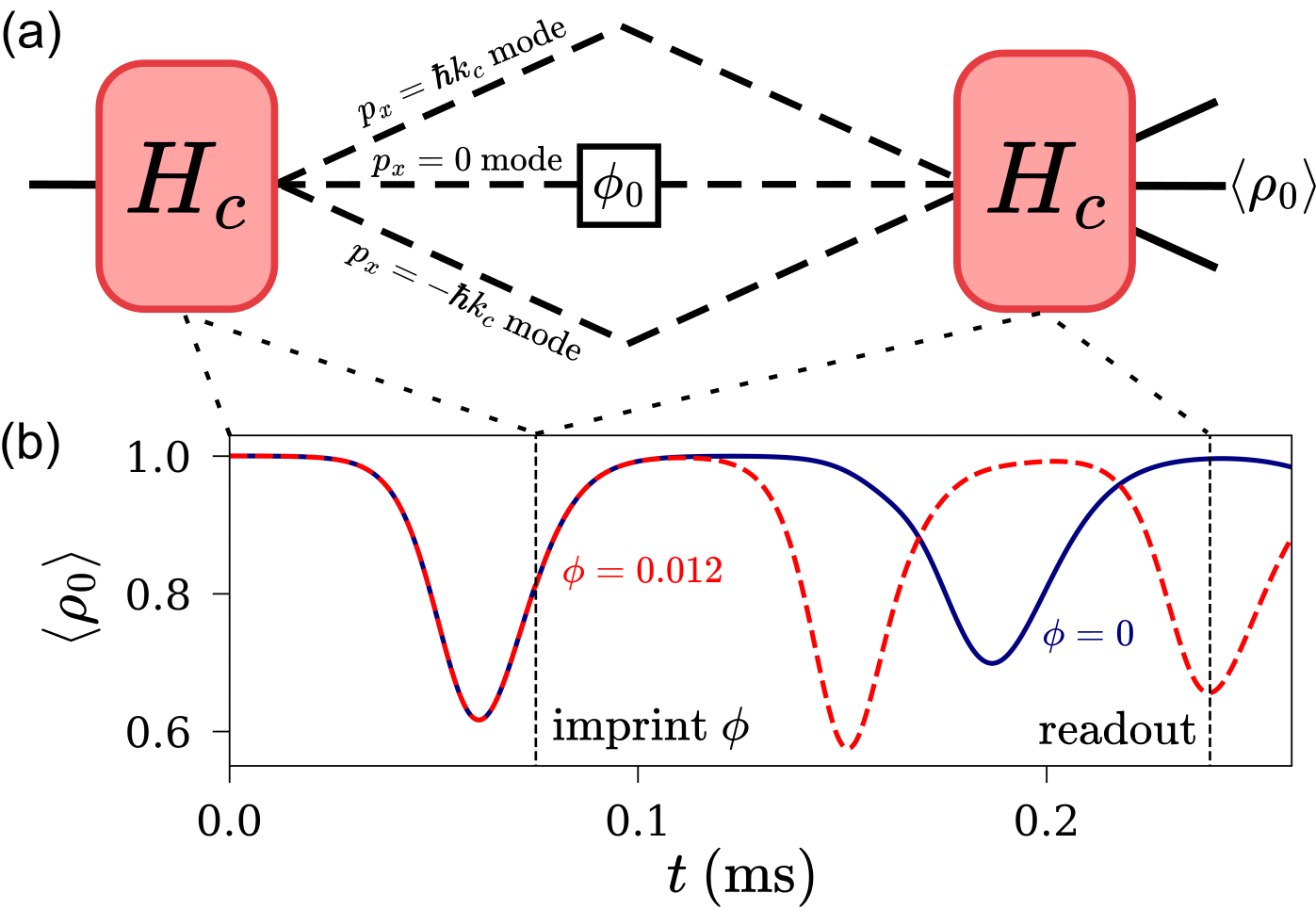}
\caption{SU(1,1) matter wave metrology using quasi-cyclic dynamics, by unitary evolution via $H_c$. (a) Principle of operation. The initial state $|N\rangle_0|0\rangle_+|0\rangle_-$ is split into the three momentum modes via unitary evolution with $H_c$, after which a phase shift $\phi=-2\phi_0=2\omega_R\tau$ is imprinted on the zero-order mode at $t=t_1$, where $\tau$ is the short time for which $\eta=0$ (see text). The quasi-cyclic evolution leads to near return to the initial state for $\phi=0$, and a phase-dependent state for $\phi>0$, at $t=t_2$. At the end of the cycle, the population in the zero-oder mode can be measured via absorption imaging in the momentum space. (b) Unitary evolution of $\langle\rho_0\rangle$ for $\phi=0$ (blue, solid) and $\phi=0.012$ (red, dashed). Vertical dashed lines indicate the pump time $t_1$ and measurement time $t_2$ (see text). Parameters: $N=10000$, $\eta=1.4\eta_c$, $\omega_R=2\pi\times 14.5$ kHz, $\bar{\Delta}_c=-1$ GHz.}\label{Fig:2}
\end{figure} 

The generation of momentum correlated pairs of atoms via $H_c$ can be explained by the process illustrated in Fig. \ref{Fig:1}b). Above the self-organization threshold, the scattering of photons into the counterclockwise (clockwise) \cite{dowling91}, initially unpopulated, ring cavity mode, leads to an atom receiving a momentum kick of $-\hbar k_c$ ($\hbar k_c$) along the $x$-axis. When this photon is scattered back into the driving field $\eta$, provided that it has not decayed out of the cavity, another atom receives a momentum kick of $\hbar k_c$ ($-\hbar k_c$) along the $x$-axis. As the same photon scatters off this atomic pair, the atoms are quantum correlated, which can lead to the appearance of momentum entangled Dicke squeezed states with reduced variance of $N_+-N_-$ \cite{dunningham2002,lucke_twin_2011,bucker_twin-atom_2011,duan_entanglement_2011,zhang_quantum_2014,lucke_detecting_2014,pezze_quantum_2018}, described in the SU(2) algebra of two modes with $p_x=\pm\hbar k_c$. Note that such states are used in linear interferometry, whereas for SU(1,1) interferometry the squeezing is best described in the three mode SU(3) algebra \cite{hamley12}. In this case the squeezing of the polar state, achieved via nonlinear pendulum-like quantum dynamics, leads to sensitivity to external perturbations \cite{gerving12,liu22}. 

Fig. \ref{Fig:1}c) depicts the maximal $\langle \rho_++\rho_-\rangle=\langle N_++N_-\rangle/N$ reached during unitary evolution for a duration of $20/\omega_R$. Due to the vanishing commutator $[N_+-N_-,H_c]=0$, the unitary evolution ($\kappa=0$) is numerically tractable by exact diagonalization even for large $N$ values \cite{law98}. Note that we here take $N$ to be conserved due to the relatively short timescales of system evolution as compared to \cite{liu22}. Below threshold, the system stays in the zero-order mode. At $\eta>\eta_c$, a macroscopic population starts appearing in the $p_x=\pm\hbar k_c$ states, which is a signature of atomic momentum ordering.

The typical unitary evolution of $\rho_0=N_0/N$ expectation values is given by the solid line in Fig. \ref{Fig:2}b). The $\langle\rho_0\rangle$ performs quasi-cyclic oscillations. Such behavior is a signature of many-body nonlinear wave mixing, and has been studied using spin models similar to $H_c$ in Refs. \cite{law98,pu99,chang05,zhang05,hamley12,gerving12}. The problem can be viewed as a nonlinear pendulum in the semiclassical treatment \cite{gerving12}. In the context of optomechanical pattern formation, the quasi-oscillations of $\langle\rho_0\rangle$ indicate sloshing dynamics, stemming from the atoms falling into and out of the optical potential wells of the self-organized lattice. For thermal atoms in the semiclassical limit, this behavior was for short timescales modeled by the Kuramoto model of coupled oscillators \cite{tesio_self-organization_2014}. 

Within the quantum description, the system starting in the polar state $|N\rangle_0|0\rangle_+|0\rangle_-$ transiently evolves to a highly squeezed state during dynamics via $H_c$, in which the system is highly sensitive to perturbations from the environment \cite{hamley12,gerving12,liu22}. Applying a small phase shift to such a state can lead to a significantly change in the final state reached at $t=t_2$, see Fig. \ref{Fig:2}b). In contrast to the rather slow evolution on the timescale of 100 ms, observed in spin-1 condensates interacting by direct interatomic collisions, here the evolution takes place on much shorter timescales of $2\pi/\omega_R\sim 100$ $\mu$s (for $\omega_R=2\pi\times 14.5$ kHz).

Self-organization via $H_c$ can also be viewed as an atomic momentum parametric amplifier, see Fig. \ref{Fig:2}a). After evolution under $H_c$ for a variable time $t_1$, a relative phase shift of $\phi=\phi_++\phi_--2\phi_0$ can be imprinted on the three momentum states \cite{liu22}. In our case a phase shift of $\phi=-2\phi_0=2\omega_R\tau$ is imprinted onto the atoms by rapidly switching off the pump laser to supress the wave mixing dynamics, and letting the system evolve via $H_c$ with $g_c=0$ for a short time $\tau$, see Fig. \ref{Fig:2}b). The switch off time for the laser is on the order of a few nanoseconds, and the intracavity photons take a time $\sim 1/\kappa$ to decay out of the cavity. For $\kappa$ values $\kappa \lesssim 5\:\omega_R$, the decay may lead to noticeable effects on the atom dynamics. However, it was shown that switching off the drive field at an appropriate time can lead to atoms reaching the desired motional state even for such small $\kappa$ values \cite{kresic2023}. The laser switch off dynamics is here approximated as an instantaneous quench of the Hamiltonian, and the optimal switch-off sequences for populating the desired atomic momentum states at $t=t_1$ will be studied in future work.

\begin{figure}[!t]
\centering
\includegraphics[clip,width=\columnwidth]{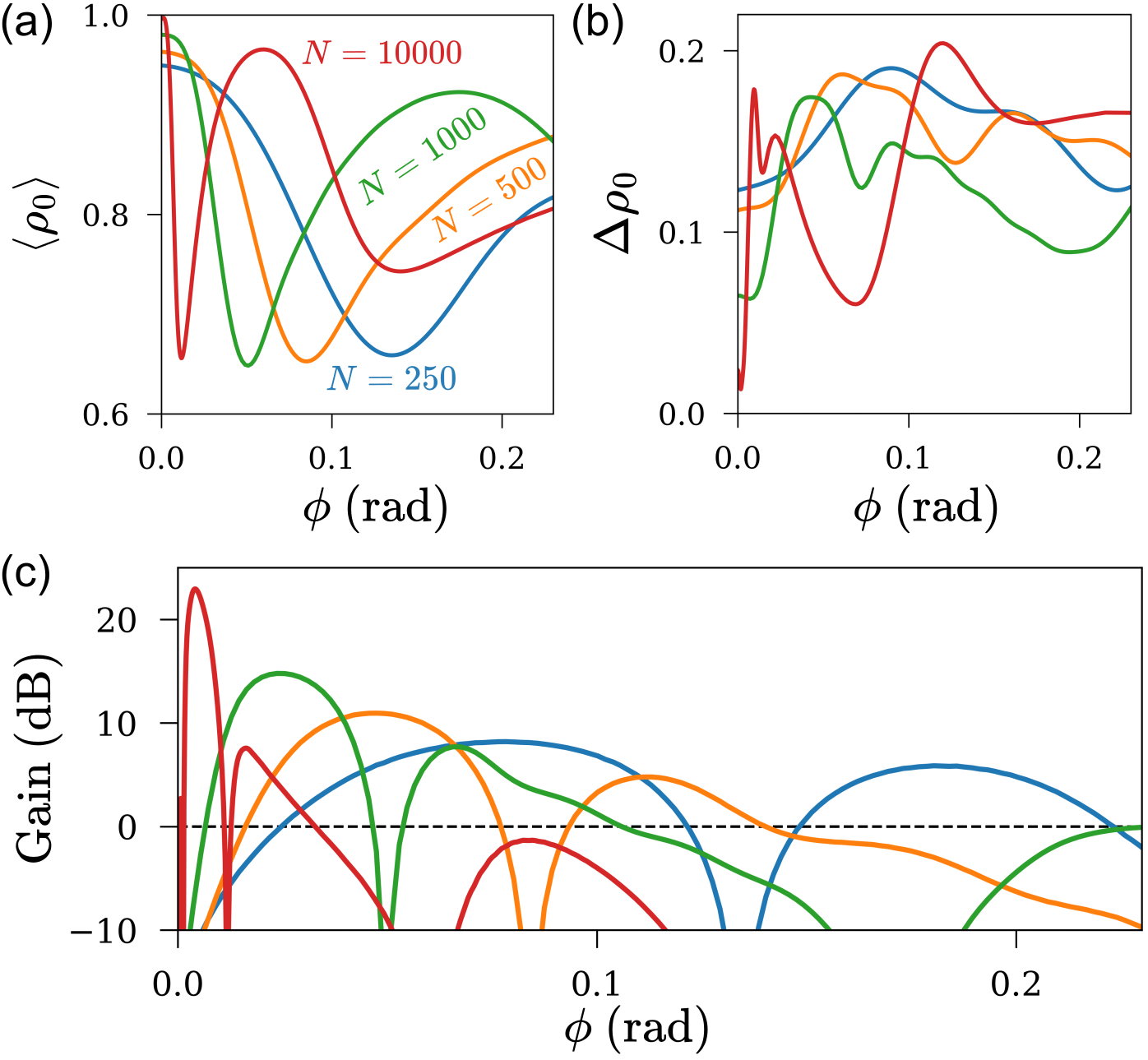}
\caption{Quantum enhancement of phase $\phi$ measurements using the quasi-cyclic evolution method illustrated in Fig. \ref{Fig:2}a). (a) $\langle\rho_0\rangle$ and (b) $\Delta\rho_0$ dependence on the imprinted phase $\phi$, for $N=250$ (blue), $N=500$ (orange) $N=1000$ (green) and $N=10000$ (red). (c) Quantum metrological gain for the same simulations, given by Eq. (\ref{eq:gain}). The horizontal dashed line indicates the standard quantum limit. Parameters: $\eta=1.4\eta_c$, $\omega_R=2\pi\times 14.5$ kHz, $\bar{\Delta}_c=-1$ GHz.}\label{Fig:3}
\end{figure}



Due to quasi-cyclic dynamics, the system for $\phi=0$ returns to approximately the initial state $|N\rangle_0|0\rangle_+|0\rangle_-$ at some time $t=t_2$. Measuring the proportion of atoms in the zero-order mode $\langle\rho_0\rangle$ and the variance thereof, via absorption imaging in momentum space, allows one to determine the value of the phase shift $\phi$. Using the value of $\tau$, which is in typical experiments known to a high degree of precision, the value of $\omega_R$ can be determined from $\phi$.

For atom numbers up to $N=500$, we here use the Schrödinger equation with a time-dependent Hamiltonian to simulate the system evolution, whereas for higher atom numbers exact diagonalization is used. In the latter case, the phase shift is imprinted by acting on the system with an operator $U_p=e^{i\phi N_0/2}$ at $t=t_1$.

The phase sensitivity of the SU(1,1) interferometer is given by the error propagation formula \cite{liu22}:
\begin{align}\label{eq:sensitivity}
\Delta\phi=\frac{\Delta\rho_0}{\left|\frac{d\langle\rho_0\rangle}{d\phi}\right|}.
\end{align}
The quantum metrological gain is given by:
\begin{align}\label{eq:gain}
\mbox{Gain}=-20\log\left(\frac{\Delta\phi}{\Delta\phi_{SQL}}\right),
\end{align}
where $\Delta\phi_{SQL}=2/\sqrt{N}$ is the phase sensitivity in the standard quantum limit, derived e.g. in \cite{liu22}. 

\begin{figure}[!t]
\centering
\includegraphics[clip,width=\columnwidth]{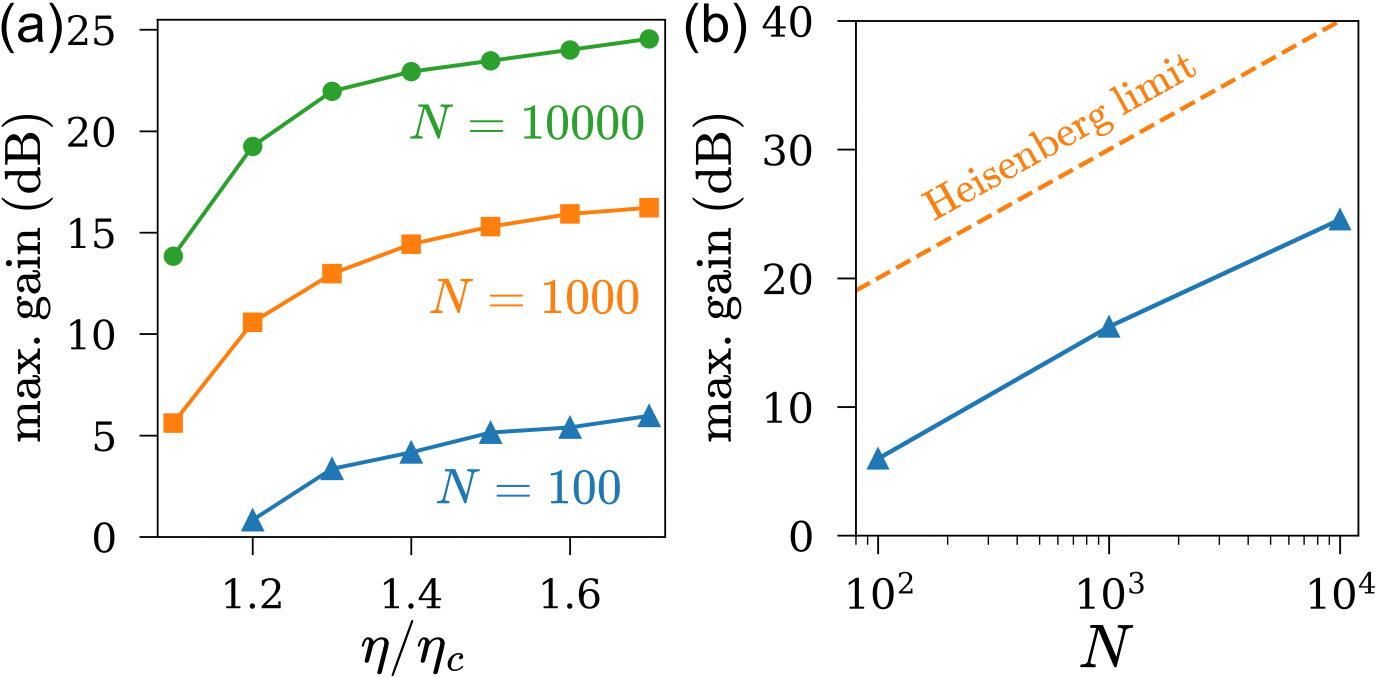}
\caption{Scaling of the maximal achieved gain with (a) pump rate $\eta$ and (b) total atom number $N$. (a) $N=100$ (blue triangles), $N=1000$ (orange squares) and $N=10000$ (green dots). (b) Numerical data (blue triangles) and the Heisenberg limit (orange dashed line), given by $20\log\sqrt{N}$. Solid lines are guide to the eyes. Parameters: $\omega_R=2\pi\times 14.5$ kHz, $\bar{\Delta}_c=-1$ GHz.}\label{Fig:4}
\end{figure}

The comparison of measurement sensitivities for $N=250,\; 500,\;1000,\;10000$ is shown in Fig. \ref{Fig:3}. For each $N$ and $\eta$ (see Fig. \ref{Fig:4}), the $t_1$ is chosen at the time with largest derivative $d\langle\rho_0\rangle/dt$, while $t_2$ is taken at the second peak of the $\langle\rho_0\rangle$ quasi-oscillation cycle, see Fig. \ref{Fig:2}b). Increasing the atom number leads to an increase in maximum quantum metrological gain, due to an increase in the slope of $d\langle\rho_0\rangle/d\phi$, see Fig. \ref{Fig:3}a,b). The $\phi$ value with maximum gain gets smaller for increasing $N$. Note that for increasing $N$, the system for $\phi=0$ returns more closely to the initial state at $t=t_2$, as $\Delta\rho_0$ gets closer to 0 and $\langle\rho_0\rangle$ gets closer to unity, see Fig. \ref{Fig:3}a,b).

The scans of maximal achieved gain with respect to $\eta$ and $N$ is shown in Fig. \ref{Fig:4}a,b). Increasing the $\eta$ near and above threshold values, the maximal achieved gain initially grows. However, the growth quickly saturates, achieving the highest value of 24.6 dB, for $N=10000$ at $\eta=1.7\eta_c$. Comparing the $N$ scaling of the values at $\eta=1.7\eta_c$ with the quantum metrological gain at the Heisenberg limit of $\Delta\phi_{Heis}/\Delta\phi_{SQL}=1/\sqrt{N}$ \cite{giovannetti_quantum_2006}, the growth is approximately parallel. The largest gain shown in Fig. \ref{Fig:4}a) is comparable to the values reported in state of the art spin squeezing experiments based on photon-mediated interaction \cite{hosten_measurement_2016,pezze_quantum_2018,colombo2022time}. 

The main source of noise in the setup stems from the decay of quantum correlations arising due to photons decaying out of the cavity with a rate $\kappa$. In the regime of $|\bar{\Delta}_c|\gg\kappa$, the transient dynamics is determined more by the coherent light-matter interaction than the photonic decay \cite{finger23}. In Appendix B, we use the Lindblad master equation and Monte Carlo wave function simulations \cite{dalibard92} to demonstrate that, for the experimentally available values $\bar{\Delta}_c=-1$ GHz and $\kappa=2\pi\times 14.5$ kHz \cite{wolke12,schuster2020}, irreversible dynamics at relevant timescales is nearly indistinguishable from unitary dynamics. Increasing the $\kappa$ values further, the irreversible dynamics leads to larger deviations of $\langle\rho_0\rangle$ and $\Delta\rho_0$ from the values for the unitary case. Namely, the $\langle\rho_0\rangle$ oscillations dephase more rapidly, while $\Delta\rho_0$ does not dramatically increase but stays approximately constant. Although a detailed study of the influence of $\kappa$ on interferometer sensitivity is beyond the scope of this Article, the simulations of irreversible dynamics give an indication that quantum enhanced SU(1,1) interferometry may be achievable for large cavity detunings even in moderate to low finesse cavities.

To conclude, we have devised and numerically explored a procedure for performing SU(1,1) matter wave interferometry beyond the standard quantum limit, with self-organized atomic momentum states in a transversely pumped ring cavity. The advantage of this light-induced SU(1,1) interferometer with respect to the procedures utilizing spin-mixing interaction, see e.g. \cite{liu22}, is the orders of magnitude speed enhancement, which allows one to neglect the atom loss out of the condensate during the relevant temporal evolution. Including the excitation of higher order momentum modes and the quantum noise arising from photon decay into the picture, will lead to complex quantum dynamics, to be explored in subsequent work. Optimization of the interferometer sensitivity in various experimental conditions is a significant future challenge, which may be researched using optimal control theory \cite{doria11} or machine learning techniques \cite{guo21}. Finally, we note that our results also have implications for the recently studied situations of \cite{finger23,kresic2023}. The proposal considered in this Article has potential for realizing quantum enhanced ultracold atom SU(1,1) matter wave interferometry in state of the art ring cavity experimental setups \cite{schuster2020}.

\textit{Acknowledgements.} We thank Paul Griffin, Helmut Ritsch and Karol Gietka for helpful discussions. The work of I. K. was funded by the Austrian Science
Fund (FWF) Lise Meitner Postdoctoral Fellowship M3011 and an ESQ Discovery grant from the Austrian Academy of Sciences (ÖAW). The dynamical evolution equations were solved numerically by using the open-source framework QuantumOptics.jl \cite{kramer_quantumopticsjl_2018}. The computational results presented here have been achieved using the Vienna Scientific Cluster (VSC).

\section{Appendix A}

We start by writing the Hamiltonian for a transversely pumped ring cavity, studied in \cite{mivehvar18}, given by:
\begin{equation}\label{eq:ham1}
\begin{aligned}
H = &-\hbar\Delta_c(n_++n_-)+ \int_V d^3r\psi^\dagger(\mathbf{r}) H_{eff}^{(1)}\psi(\mathbf{r}),
\end{aligned}
\end{equation}
where $\Delta_c=\omega-\omega_c$ is the laser-cavity detuning, $n_\pm=a_\pm^\dagger a_\pm$, and the effective single-particle Hamiltonian is given by:
\begin{equation}\label{eq:ham2}
\begin{aligned}
H_{eff}^{(1)} =& \frac{\mathbf{p}^2}{2m}+ \hbar U_0(n_++n_-+a_+^\dagger a_-e^{-2ik_cx}+a_-^\dagger a_+e^{2ik_cx})\\
+&\hbar\eta(a_+e^{ik_cx}+a_-e^{-ik_cx}+\mbox{H.c.}),
\end{aligned}
\end{equation}
where $\eta=G_0\Omega/\Delta_a$ is the maximum depth of the optical potential per photon due to the scattering between pump and cavity modes (i.e. $\eta$ - cavity pump rate) and $U_0=G_0^2/\Delta_a$ is the maximum depth of the optical potential per photon due to the scattering between cavity modes, with $G_0$ being the cavity mode coupling strength, $\Omega$ the Rabi frequency and $\Delta_a=\omega-\omega_a$ the laser detuning from the atomic optical transition. 

Taking only the zeroth and first order momentum modes into account, the atomic field operator is given by:
\begin{align}\label{eq:atomfield_appendix}
\psi(\mathbf{r})=\frac{1}{\sqrt{V}}\left(b_0+b_+e^{ik_cx}+b_-e^{-ik_cx}\right),
\end{align}
where $b_j$ is the bosonic annihilation operator of the $j$-th transverse atomic momentum mode. 

We insert Eq. (\ref{eq:atomfield_appendix}) into Eq. (\ref{eq:ham1}) for a real-valued pump rate $\eta$ and perform the integration over the BEC cloud volume $V$ to get the effective total Hamiltonian $H = H_0+H_{int}$, where the noninteracting part $H_0$ has now the form ($\hbar=1$):
\begin{equation}\label{eq:h0}
\begin{aligned}
H_0 &= -\bar{\Delta}_c(n_++n_-) +\omega_R(N_++N_-),
\end{aligned}
\end{equation} 
where $\bar{\Delta}_c=\Delta_c-NU_0$, $N_\pm=b_\pm^\dagger b_\pm$, and the light-matter interaction terms are:
\begin{align}\label{eq:fwmterms}
H_{int}& =  U_0 a_+^\dagger a_-b_-^\dagger b_++
\eta(a_++a_-^\dagger)(b_+^\dagger b_0+b_0^\dagger b_-)+\mbox{H.c.}.
\end{align}
Near threshold and/or for $\Omega\gg G_0$, the term $U_0 a_+^\dagger a_-b_-^\dagger b_+$ is small compared to the terms proportional to $\eta$. Using now the Hamiltonian $H_{int}'$:
\begin{align}\label{eq:fwmterms2}
H_{int}'& = \eta(a_+^\dagger b_-^\dagger b_0+a_+b_+^\dagger b_0+a_-^\dagger b_+^\dagger b_0+a_-b_-^\dagger b_0)+\mbox{H.c.},
\end{align}
one gets for the input-output equations of the intracavity field operators $a_\pm$ \cite{gardiner_input_1985,scully01}:
\begin{align}\label{eq:heiseq_phot1}
\frac{d a_\pm}{d t}& = (i\bar{\Delta}_c-\kappa)a_\pm-i\eta(b_\mp^\dagger b_0+b_0^\dagger b_\pm)+\xi_\pm(t),
\end{align}
where $\xi_\pm(t)$ are the quantum noise operators of the cavity modes.

We now adiabatically eliminate the photonic degrees of freedom $a_\pm$ by neglecting the $\xi_\pm(t)$ terms in the above equations and setting $\dot{a}_\pm=0$. Inserting this $a_\pm$ into the Hamiltonian $H'=H_0+H_{int}'$, we get the Hamiltonian for the atomic momentum subsystem:
\begin{align}\label{eq:heiseq_phot2}
H_c& = g_c'[2b_+^\dagger b_-^\dagger b_0 b_0+2b_0^\dagger b_0^\dagger b_+ b_-\\
+& (2N_0-1)(N_++N_-)-2N_0]-\omega_RN_0,
\end{align}
where $g_c'=\bar{\Delta}_c\eta^2/(\bar{\Delta}_c^2+\kappa^2)=\omega_R \eta^2/(4N\eta_c^2)$, $\eta_c=\sqrt{-\omega_R(\bar{\Delta}_c^2+\kappa^2)/(4\bar{\Delta}_cN)}$, and we have used $N=N_0+N_++N_-$. 

Note that in deriving $H_c$ we have neglected the photonic quantum noise terms in the input-output formalism. The reasoning for this is that the photonic modes are initially in a vacuum state, and we work in the limit $|\bar{\Delta}_c|\gg\kappa$ \cite{sorensen02,finger23}, where the photon decay is expected to only weakly influence the atomic motion. 

The cavity dissipation for the adiabatically eliminated photonic modes is below included at the level of the Lindblad master equation, which describes the influence of cavity photon decay on the creation of atomic momentum pairs. For transverse patterns in a longitudinally pumped ring cavity setup, this treatment was corroborated by numerical results, and excellent agreement with experimental results was also reported for self-organization in a single mode Fabry-Perot resonator with two-level ground state atoms, exhibiting similar physics \cite{finger23,kresic2023}. 

Note also that for a single mode cavity driven longitudinally near resonance \cite{murch08}, an atomic diffusion term was shown to arise due to photonic quantum noise \cite{nagy_nonlinear_2009}. This was interpreted as a consequence of backaction on the atomic momentum, arising due to photodetection measurement of the photons leaking out of the cavity. This backaction is related to the fact that, for single mode cavities, the measurement of the number of photons leaking out of the cavity can provide information about the collective atomic position (i.e. density distribution), which is an operator conjugate to collective momentum. The analysis of the magnitude of the backaction term, and its influence on the quantum dynamics of the system, for the continuously translationally symmetric Hamiltonian $H$, is an intriguing topic for future research.


\begin{figure}[!t]
\centering
\includegraphics[clip,width=\columnwidth]{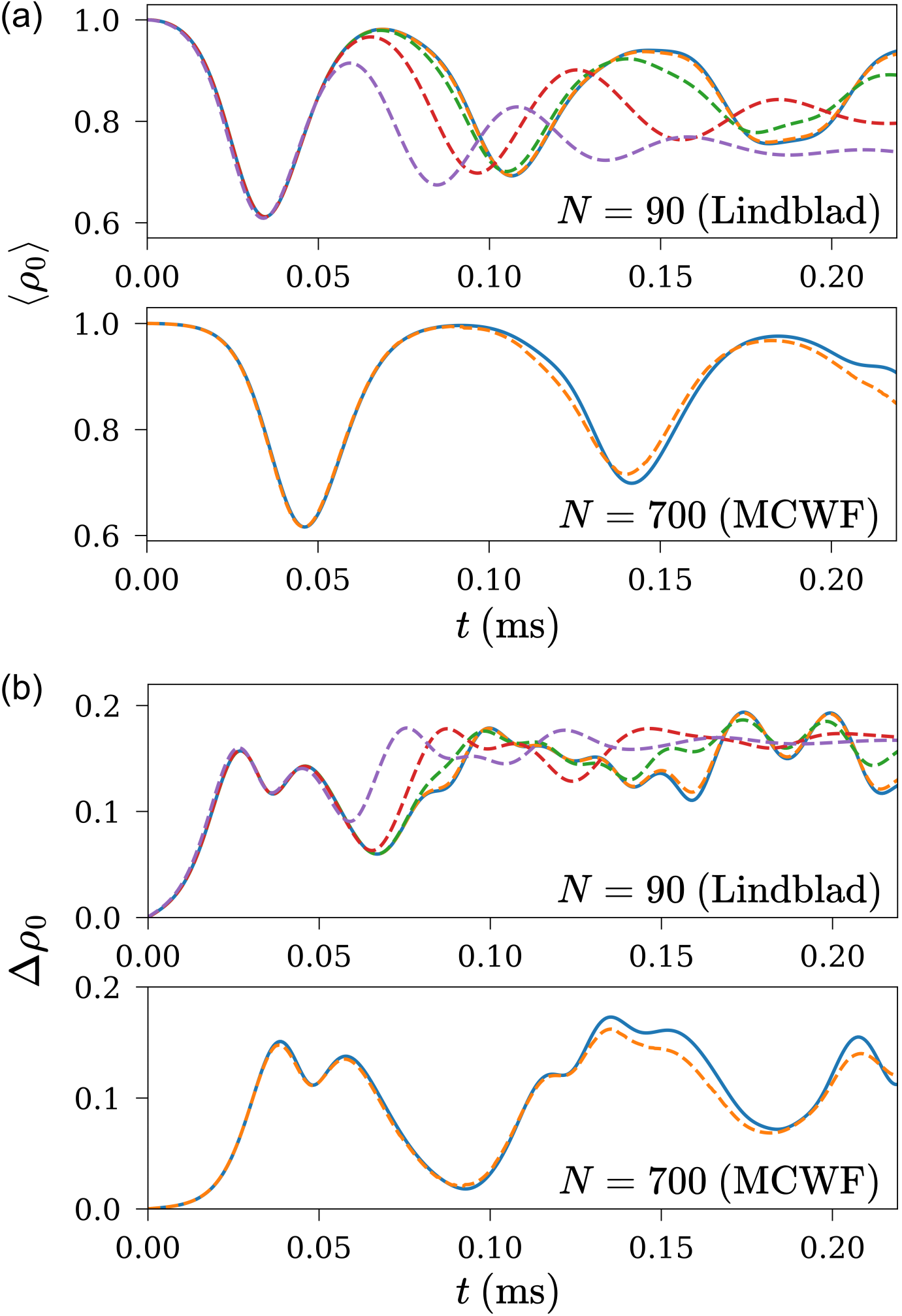}
\caption{Comparison of unitary and irreversible evolution of (a) $\langle\rho_0\rangle$ and (b) $\Delta\rho_0$. In all figures, the solid blue line is the result for unitary dynamics, while dashed lines are the results for  $\kappa/2\pi= 14.5$ kHz (orange), $145$ kHz (green), $1.45$ MHz (red), $14.5$ MHz (purple). In addition to the solutions to the Lindblad equation, we plot the results of Monte Carlo wave function (MCWF) simulations for $\kappa/2\pi=14.5$ kHz, averaged over $100$ trajectories. Parameters: $N=90$ (Lindblad), $N=700$ (MCWF), $\eta=1.4\eta_c$, $\omega_R=2\pi\times 14.5$ kHz, $\bar{\Delta}_c=-1$ GHz.}\label{Fig:5}
\end{figure} 

\section{Appendix B}
The influence of cavity photon dissipation on the evolution of atomic degrees of freedom can be described by the Lindblad equation \cite{kresic2023}:
\begin{align}\label{eq:lindrhoadiab2}
\frac{d\rho}{dt}&=-\frac{i}{\hbar}[H_c,\rho]\\
+&\gamma\sum_{j=\pm}(2K_j \rho K_j^\dagger-K_j^\dagger K_j \rho-\rho K_j^\dagger K_j), 
\end{align}
with:
\begin{align}\label{eq:newjump}
\gamma = \frac{\kappa \eta^2}{(\bar{\Delta}_c^2+\kappa^2)},\;K_\pm=(b_\mp^\dagger b_0+b_0^\dagger b_\pm),
\end{align}
describing the influence of cavity photon decay on the atomic momentum pair creation. The typical cavity dissipation rates $\kappa/(2\pi)$ in ultracold atom experiments range from values on the order of a few MHz \cite{schmidt2014,rivero22}, down to values of a few kHz \cite{wolke12,schuster2020}. Note also that free spectral ranges for commonly used cavities are on the order of a few GHz, and in our simulations we fix the detuning at $\bar{\Delta}_c=-1$ GHz. 

Along with solving the Lindblad equation, irreversible evolution of the system was studied using Monte Carlo wave function calculations \cite{dalibard92}, with jump operators $\sqrt{2\gamma}K_\pm$. The influence of experimentally realistic $\kappa$ values on the evolution of $\langle\rho_0\rangle$ and standard deviation $\Delta\rho_0=\sqrt{\langle \rho_0^2\rangle-\langle \rho_0\rangle^2}$ is shown in Fig. \ref{Fig:5}. At high finesse cavity value $\kappa/2\pi= 14.5$ kHz, the curves closely follow the ones of the unitary evolving case. For increasing the $\kappa$ further, more noticeable deviations from the unitary case are observed. For $\langle\rho_0\rangle$, the oscillations start going out of phase from the unitary case, with the oscillation amplitude reducing for longer times at larger $\kappa$'s. The $\Delta\rho_0$ does not dramatically increase for increasing $\kappa$, which is a promising indication for potential experimental realizations.

\bibliography{references}
\clearpage
\newpage

\end{document}